\documentclass[12pt]{article}
\pdfoutput=1
\usepackage{epsfig}
\usepackage{amsfonts}
\usepackage{amssymb}
\usepackage{xcolor}
\usepackage{amsmath}
\usepackage{upgreek}
\usepackage{graphicx}
\usepackage{cite}
\usepackage{calligra}
\usepackage{caption}
\usepackage{subcaption}
\usepackage{cite}
\usepackage{calligra}
\usepackage{caption}

\newcommand{\nc}{\newcommand}
\nc{\beq}{\begin{equation}} \nc{\eeq}{\end{equation}}
\nc{\beqa}{\begin{eqnarray}} \nc{\eeqa}{\end{eqnarray}}
\nc{\ba}{\begin{array}} \nc{\ea}{\end{array}}

\begin{document}
\begin{center}

{\bf \LARGE Three-loop chiral effective potential\\[0.3cm] in  the Wess-Zumino model} \vspace{1.0cm}

{\bf \large I.L. Buchbinder$^{1,2,3,a}$, R.M. Iakhibbaev$^{1,b}$, D.I. Kazakov$^{1,c}$ \\[0.3cm] and D. M. Tolkachev$^{1,4,d}$}

\vspace{0.5cm}
{\it $^1$Bogoliubov Laboratory of Theoretical Physics, Joint Institute for Nuclear Research,
  6, Joliot Curie, 141980 Dubna, Russia\\
$^2$ Department of Theoretical Physics, Tomsk State Pedagogical University,
634041 Tomsk, Russia \\
$^3$ National Research Tomsk State University,
Lenin Av. 36, 634050 Tomsk, Russia \\ and \\
$^4$Stepanov Institute of Physics,
68, Nezavisimosti Ave., 220072, Minsk, Belarus\\}
\vspace{0.5cm}

\abstract{We calculate the three-loop contribution to the chiral effective
potential in the massless Wess-Zumino model. It is shown that while the non-renormalisation  theorem forbids divergent contributions to the chiral potential, in the massless case the finite corrections survive. There are only three three-loop supergraphs that give rise to a superfield effective action in the pure chiral sector. Two of them are UV finite while the third requires one-loop counterterm corresponding to the chiral field renormalisation.}

\textit{Keywords}: {$\mathcal{N}=1$ supersymmetric field theory, chiral superfields, Wess-Zumino model,
quantum effective action}
\end{center}
\text{\footnotesize{E-mails: $^a$buchbinder@theor.jinr.ru, $^b$yaxibbaev@jinr.ru,
$^c$kazakovd@theor.jinr.ru, $^d$den3.1415@gmail.com}}

\section{Introduction}
Supersymmetric quantum field theory presents various surprises related to an unexpected cancelation of formally admissible contributions and can sometimes even lead to exact results (see e.g.\cite{Seiberg-1}, \cite{Terning} and the references therein). Many of these surprises in theories formulated in ${\mathcal N}=1$
superspace with coordinates $(x^m,\,\theta^{\alpha},\,\bar{\theta}^{\dot{\alpha}})$ result from the non-renormalisation theorem (see e.g. \cite{GGRS,BK}).
This theorem states that any contribution to the effective action can be written as a single integral over $d^4\theta$ \footnote{${\mathcal
N}=2$ supersymmetric theories can be formulated in terms of
unconstrained superfields in harmonic superspace \cite{GIOS}. The corresponding ${\mathcal N}=2$ non-renormalisation theorem was given in \cite{BKO} and is analogous in some aspects to the ${\mathcal N}=1$ non-renormalisation theorem.}.
The non-renormalisation theorem allows one to describe the  general renormalisation structure of an arbitrary ${\mathcal N}=1$ supersymmetric theory in a simple and universal manner. A method of evaluating Feynman diagrams, which in this context are usually called supergraphs, related to the non-renormalisation theorem opens up the possibility of finding finite contributions to the effective action in superfield terms (see e.g., \cite{West-1, Jack, Dunbar, West-2, Buchbinder:1993ud, Buchbinder:1994iw,
 Buch, West-3, BCP, BCP-1, BP, Kuz, Martin:2024qmi} and references therein).

As a consequence of the non-renormalisation theorem, one can expect that the chiral potential in ${\mathcal N}=1$ supersymmetric theories should get neither infinite nor finite quantum corrections. This is indeed true for infinite corrections and finite corrections in massive theories. However, massless theories present another
surprise. In Ref.\cite{West-1}, it was noted and then confirmed by direct component \cite{Jack} and superfield \cite{Buch} calculations that the chiral potential in the Wess-Zumino model acquires at least a
two-loop finite correction \footnote{In essence, this correction means an anomaly of Seiberg's holomorphy principle \cite{Seiberg-2} and is inherent only in the one-particle irreducible effective action in massless theories}. Later it was shown that such a finite two-loop correction to the chiral effective potential appears even in the non-renormalisable (anti)chiral theory with arbitrary Kähler and
chiral potentials \cite{BP}, which is one more surprise.

Natural questions arise in this regard as to gow to interprent all these surprises. How does the existence of finite contributions to the chiral effective potential for massless theories relate to the general statement of the non-renormalisation theorem, according to
which any contribution to the effective action can be written as an integral over a general superspace? Why does the effect in question exist only in massless theories?  What happens in higher loops?

In the proposed paper, we will answer these questions for the Wess-Zumino model and present explicit calculations in the
three-loop approximation.

The answers to the  mentioned questions are based on
the identity for the chiral superfield 
\beq 
\label{identity} 
\int d^8 z ~ u(\Phi) \left(-\frac{D^2}{4 \square}\right) v(\Phi)=\int d^6z ~u(\Phi) v(\Phi).
\eeq 
Although the integral in the left-hand side is written as an integral over full superspace, it is identically
transformed into an integral over chiral subspace due to the non-local operator $-\frac{D^2}{4 \square}$. 
The operator $\frac{1}{\square}$ can be originated in supergraphs only from massless propagators. Moreover, qualitatively one can expect that the final contributions to the effective action stipulated by the identity \eqref{identity} should be finite since the divergent contributions are local in space-time \footnote{To be more precise, this point needs a more careful discussion. Let us consider the $L$-loop contribution to the effective action in terms of bare fields and coupling constants, and let as a result of calculations we arrive at the expression in full superspace containing identity \eqref{identity}. This expression must be finite since the non-local
operator $\frac{1}{\square}$ can be originated only as a  divergence in full superspace that must be space-time local. However, the coefficients in the above expression can be divergent due to possible divergences of subdiagrams corresponding to loops with numbers less then given $L$.}. Therefore, the final chiral quantum contributions to the effective action are finite ( after the renormalisation of the fields) and can exist only in massless theories. In the
proposed paper, we demonstrate this phenomenon for the case of the massless  Wess-Zumino model  in the three-loop approximation.

We consider the four-dimensional Wess-Zumino model  with the action
\beq
\label{WZact} \mathcal{S}[\Phi,\bar{\Phi}]=\int d^8z ~\Phi
\Bar{\Phi}+ \frac{\lambda}{3!}\int d^6z~\Phi^3+ \frac{\bar{\lambda}}{3!}\int d^6\bar{z}~\bar{\Phi}^3,
\eeq
where $\Phi$ is the chiral superfield, $\bar{\Phi}$  is the
anti-chiral superfield, and $\lambda$ and $\bar{\lambda}$ are the complex conjugate coupling constants.
The integral in the first term is taken over full superspace with the measure
$d^8z=d^4x d^2\theta d^2\Bar{\theta}$, the integral in the second term is taken over chiral subspace with the
measure $d^6z=d^4x d^2\theta$, and the integral in the last term in the action is taken over antichiral
subspace with the measure $d^6\bar{z}=d^4x d^2\bar{\theta}$. The purpose of the paper is to calculate three-loop quantum corrections to the classical
chiral potential $W(\Phi)= \frac{\lambda}{3!}\Phi^3$.

The paper is organized as follows. In Section 2 we discuss a procedure for calculating a superfield
effective action in the model under consideration. Section 3 is devoted to a brief discussion of the known
two-loop chiral correction. In Section 3 we consider calculations of the chiral effective potential
at three loops. There are three three-loop superfield Feynman diagrams. Two of them are finite
and one requires a one-loop counterterm to eliminate the divergence in the propagator-type  subdiagram.
Section 4 contains the  summary of the results.

\section{Superfield  effecive potential}

The superfield effective action $\Gamma[\Phi, \bar{\Phi}]$ is defined as Legedre transform of generating functional of the
one-particle irreducible (1PI) Green functions and is given in terms of the following functional integral (see e.g. \cite{BS}):
\begin{eqnarray}
\label{EA}
 \exp\left(\frac{i}{\hbar}\Gamma[\Phi,\bar{\Phi}]\right) =
 \int {\cal D} \phi {\cal D} \bar{\phi}~
\exp\frac{i}{\hbar} \mathcal{S}[\Phi+\sqrt{\hbar}\phi,\bar{\Phi}+\sqrt{\hbar}\bar{\phi}] \nonumber\\
-
 \left(\sqrt{\hbar} \int d^6z ~
  \phi(z) \frac{\delta{\Gamma}}{\delta\Phi(z)} + h.c.\right)]
\end{eqnarray}
Relation (\ref{EA}) defines a loop expanding the effective action. The effective action can be written as a sum of classical action and quantum corrections,
$\Gamma[{\Phi}, \bar{\Phi}]=
\mathcal{S}[\Phi, \bar{\Phi}] + \bar{\Gamma}[\Phi, \bar{\Phi}]$, 
where $\bar{\Gamma}[\Phi, \bar{\Phi}]$
contains all the quantum corrections and can be expanded in power series in $\hbar$. The terms of
this expansion are obtained by expanding the right-hand side of (\ref{EA}) in $\hbar$. In this context,
the superfields $\phi,\, \bar{\phi}$ in the functional integral are called quantum and the superfields $\Phi,\,\bar{\Phi}$ on which the effective action depends are called background. The result of the expansion is expressed in terms of supergraphs where the quadratic part of classical action on the background of superfields $\Phi,\, \bar{\Phi}$ defines the background superfield-dependent propagators and the part cubic in quantum fields of the classical action defines the vertices.

The quadratic part of the Wess-Zumino action (\ref{WZact}) in the superfields $\Phi,\, \bar{\Phi}$ has the form
\begin{eqnarray}
\label{quadract}
\mathcal{S}^{(2)}[\phi,\bar{\phi}] & = & \int d^8z~\phi\bar{\phi} +
   \left(\frac{1}{2}\int d^6z~ \lambda \Phi \phi^2 + h.c.\right).
\end{eqnarray}
The corresponding super-propagator is the matrix with the elements $G_{ab}$ where $a,b\,=+,-$ and given by
\begin{equation}
\label{def}
 \left(
 \begin{array}{cc}
  -\lambda \Phi & \frac{1}{4}\bar{D}^2\\
  \frac{1}{4}D^2 &  -\lambda \bar{\Phi}
 \end{array}
 \right)
 \left(
 \begin{array}{ll}
  G_{++} & G_{+-}\\
  G_{-+} & G_{--}
 \end{array}
 \right)
 =- \left(
 \begin{array}{ll}
  \delta_+ & 0\\
  0 & \delta_-
 \end{array}
 \right)
\end{equation}
where $\delta_+=-\frac{1}{4}\bar{D}^2\delta^8(z_1-z_2)$,
$\delta_-=-\frac{1}{4}D^2\delta^8(z_1-z_2)$ are the chiral and antichiral superspace $\delta$-functions, respectively.  The signs $+$ and $-$ of the matrix elements  of the super-propagator  denote the chirality or anti-chirality in the corresponding argument (see the details in \cite{BK}).

In the case when the background superfields $\Phi$ and $\bar{\Phi}$ satisfy the supersymmetric conditions
\begin{equation}\label{conditions}
 \partial_a\Phi = \partial_a\bar{\Phi} = 0,
\end{equation}
the superfield effective action is expressed in terms of the superfield effective potentials in the form
\beq
\label{sep}
\Gamma[\Phi,\bar{\Phi}]= \int d^8z~ \left(\textbf{K}(\Phi,\,\bar{\Phi}) +
\textbf{A}(D\Phi,\bar{D}\bar{\Phi})\right) \nonumber \\
+ \left(\int d^6z~ \textbf{W}(\Phi) + c.c.\right)\, .
\eeq
where $\textbf{K}(\Phi,\,\bar{\Phi})$ is the Kähler effective potential, $\textbf{A}(D\Phi,\bar{D}\bar{\Phi})$ is the auxiliary fields effective potential and $\textbf{W}(\Phi)$ is the chiral effective potential which is the main focus of this paper.
In the framework of loop expansion, the superfield
effective potential can be written as a series in the number of loops
\beq
\label{loopexp}
\Gamma[\Phi,\bar{\Phi}]=\sum_{L=1}^\infty\hbar^L \Gamma^{(L)}[\Phi,\bar{\Phi}].
\eeq
Therefore, each of the effective potentials $\textbf{K},\, \textbf{A}$ and $\textbf{W}$ can also
be written in the form (\ref{loopexp}).

Since we are interested in calculating the chiral effective
potential, we need to find a solution to equation (\ref{def})
for the background superfields satisfying condition
(\ref{conditions}), and it is sufficient to put $\bar{\Phi}=0.$ This solution was found in Ref. \cite{Buch}
\beq \label{G}
\mathcal{G}=\mathcal{G}_{0} + \mathcal{G}_{1},
\eeq
where
\beq
\label{G0} \mathcal{G}_0 = - \frac{1}{16}\begin{pmatrix} 0  &
\frac{\bar{D}_{1}^{2}D_{2}^{2}}{\Box_1}
\\ \frac{D_{1}^{2}\bar{D}_{2}^{2}}{\Box_1} & 0 \end{pmatrix} \delta^{8}(z_1-z_2)  \,
\eeq
is free super-propagator and
\beq
\mathcal{G}_1 = - \frac{1}{16}\begin{pmatrix} 0  & 0
\\ 0 & \mathcal{G}(z_1,z_2) \end{pmatrix}   \,
\label{G1}
\eeq
is the background superfield-dependent correction with
\beq \label{propagator}
\mathcal{G}(z_1,z_2)=-\frac{D_{1}^{2}D_{2}^{2}}{\Box_1}(\lambda\Phi_1\frac{\bar{D}_{1}^{2}}{4\Box_1}
\delta^{8}(z_1-z_2)).
\eeq
Note that the function $\mathcal{G}(z_1,z_2)$ is anti-chiral in each of its arguments. Namely, the super-propagator (\ref{G}-\ref{propagator}) should be used in loop
calculations of the chiral effective potential.

The possibility of obtaining chiral loop corrections to the classical potential was analyzed in Ref.
\cite{BP} where it was shown that such corrections are possible under the conditions
\beq
2L+1-n_2-n_3=0,\; ~ n_{D^2}+1=n_{\bar{D}^2}, \label{cond}
\eeq
where $L$ denotes the loop number and $n_2$ means the number of vertices of the form $\Phi\phi^2$. Since
all possible chiral corrections should be proportional to $\Phi^3$, one gets $n_2 = 3.$ The $n_3$
means a number of vertices of the form $\phi^3$ (there are no higher order derivatives in the  Wess-Zumino
model) in the supergraph, $n_{D^2}$ and $n_{\bar{D}^2}$ are the numbers of the corresponding supercovariant
derivatives in the supergraph. Note that the vertices with all convolved lines have two covariant derivatives, whose form depends on the chirality or anti-chirality of this vertex. If the condition \eqref{cond} is not
satisfied, the contribution vanishes or leads to the infrared pole. Relations (\ref{cond}) show that the chiral contributions begin with a two-loop approximation.
Note that the second of the conditions \eqref{cond} arises from the dimensional consideration \cite{BP} and identity \eqref{identity}.

\section{Discussing two-loop chiral correction}

\begin{figure}[ht]
 \begin{center}
  \epsfxsize=4.0cm
 \epsffile{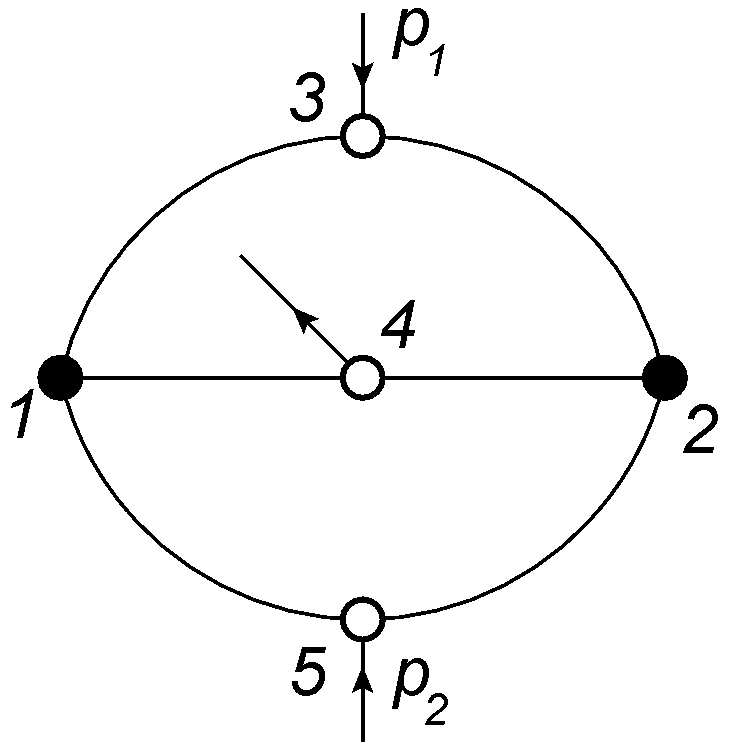}
 \end{center}
 \vspace{-0.2cm}
 \caption{Graphs contributing to the holomorphic effective potential at the two-loop level. White blobs are antichiral vertices, black ones are chiral vertices, and $p_1$ and $p_2$ are incoming momenta.}
\label{2loop}
 \end{figure}

The two-loop contribution to the effective chiral potential is
calculated as follows. In the theory under consideration, the
two-loop effective action is described by the only supergraph
(walnut-like diagram) without external lines, where the dependence on the background field
is contained in the propagators, i.e., in the internal lines. When substituting the propagator $\mathcal{G}=\mathcal{G}_0 +
\mathcal{G}_1$ with $\mathcal{G}_0$ (\ref{G0}) and $\mathcal{G}_1$ (\ref{G1}) into this supergraph, one finally arrives at the supergraph in Fig. \ref{2loop}. The corresponding contributions can be written as
\cite{Buch}
\beq \Gamma^{(2)}(\Phi)=\frac{\bar{\lambda}^2}{3!\times
2}\int d^6\bar{z}_1 d^6 \bar{z}_2 ~ (\mathcal{G}(z_1,z_2))^3,
\eeq
where $\mathcal{G}(z_1,z_2)$ is given by (\ref{propagator}). It
should be noted that such an integral can be converted into a standard supergraph that should be calculated under the condition that external momenta are zero (this is a definition of the effective potential which is part of the effective action without derivatives).
This supergraph can be rewritten using supergraph manipulation techniques as pointed in the previous section
\beq
\begin{gathered}
\Gamma^{(2),h}(\Phi)=\lim_{p_1,p_2\rightarrow0}\frac{\lambda^5}{12}\int \prod_{i=1}^5 d^8 z_i \Phi(z_3)\Phi(z_4)\Phi(z_5) \left\{ \frac{1}{\square_1}\delta_{1,3} \frac{D^2_2 \bar{D}^2_3}{16\square_2} \delta_{3,2}  \right. \\ \left. \frac{1}{16 \square_2} \delta_{2,4} \frac{D^2_1\bar{D}^2_4}{16\square_1} \delta_{1,4} \frac{D^2_1\bar{D}^2_5}{16\square_1} \delta_{1,5}\frac{D^2_2}{4\square_2}\delta_{2,5} \right\}
\end{gathered}\label{twoloopDint}
\eeq
where $\delta_{i,j}=\delta^8(z_i-z_j)$.

The first problem we have to overcome is to perform complicate enough $D$-algebra computations that should eventually
turn the supergraph contributions into scalar momentum master integrals. These calculations can be
efficiently and simply enough fulfilled using the Wolfram Mathematica packages (see e.g. \texttt{SusyMath.m} \cite{Ferrari:2007sc}). The scalar master-integrals themselves can be evaluated in numerous of ways , for example, by the uniqueness relation or the Mellin-Barnes representation method
\cite{Kazakov:1983pk,Chicherin:2012yn,Grozin:2012xi}. If we use the uniqueness method, for example,
we can see that the square of momentum ($q_1^2$ or $q_2^2$) in the numerator reduces one of the
squares of momentum in the denominator. So it becomes possible to apply the star-triangle relation and integration by parts to one of the loops and the calculation of such a diagram becomes very straightforward
\cite{Kazakov:1983pk,Kazakov:1984mm}.

Expression \eqref{twoloopDint} after the $D$-algebra routine can be reduced to
\beq
\begin{gathered}
I_2(p_1,p_2)=\lim_{p_{1,2}\rightarrow0}\int \frac{d^4 q_1}{(4\pi)^4} \frac{d^4 q_2}{(4\pi)^4}  \frac{(q_1^2 p_1^2+q_2^2 p_2^2 -2(q_1 \cdot q_2)(p_1 \cdot p_2))}{q_1^2q_2^2 (q_1+q_2)^2 (q_1-p_1)^2 (q_2-p_2)^2 (q_1+q_2-p_1-p_2)^2}
\label{twoloopmaster}
\end{gathered}
\eeq
The loop integrals should be calculated using dimensional regularisation $d=4-2\epsilon$ and expansion by the regulator $\epsilon$ and then obtained in the limit assuming $p_{1}, p_{2} \rightarrow 0$ (see Appendix \ref{Appendix}).

Under the assumption that the superfields are slowly changing $$\Phi(y_1,\theta) \Phi(y_2,\theta) \Phi(x,\theta)\simeq \Phi^3(x,\theta)$$ the result in superfield form was obtained in Ref.\cite{Buch}\footnote{In the papers it was considered in the case when the coupling constant is real, $\lambda=\bar{\lambda}$. To show that the result (\ref{W2}) means Seiberg holomorphy anomaly, we specially
recalculated the two-loop chiral contribution for the case where the coupling constants in the classical chiral
and anti-chiral potentials are complex and conjugate one to another. The $\textbf{W}^{(2)}$ (\ref{W2})
contains both $\lambda$ and $\bar{\lambda}$ which violates the Seiberg holomorphy principle \cite{Seiberg-2}
where the chiral effective potential should depend only on $\lambda$ but be independent of
$\bar{\lambda}$.}
\beq
\label{W2}
\textbf{W}^{(2)}=\frac{|\lambda|^4}{(4\pi)^4} \frac{1}{2}\zeta(3)~\lambda\Phi^3,
\eeq
where $\zeta(n)$ is the Riemann zeta-function.

Note that the result (\ref{W2}) is finite and obtained in terms of bare superfield and coupling
constants. No divergences appeared during the calculations and no renormalisation is required.
However, it is well known that the Wess-Zumino model as a whole is a renormalisable quantum field theory
(see e.g., \cite{GGRS,BK}), where the bare and renormalised quantities are related to one  another as follows:
\beqa
&& \Phi = z^{\frac{1}{2}}{\Phi}_R,\,\, \bar{\Phi} =
z^{\frac{1}{2}}{\bar{\Phi}}_R, \label{renorm}\\
&&\lambda = z^{-\frac{3}{2}}\lambda_R,\,\,
\bar{\lambda} = z^{-\frac{3}{2}}\bar{\lambda}_R
\eeqa
with $z=1 +o(\hbar)$ being the only renormalisation constant whose expansion starts with $|\lambda|^2$.

It is remarkable that up to two loops the chiral effective potential can be equally written in terms of bare or renormalised fields and couplings. Indeed one has in two loops
\beq
W(\Phi)=\lambda \Phi^3+\frac{|\lambda|^4}{(4\pi)^4} \frac{1}{2}\zeta(3)~\lambda\Phi^3+O(|\lambda|^6)\lambda\Phi^3, \label{twoloop}
\eeq
and, according to the renormalisation rules written above, the product $\lambda \Phi^3$ is not renormalised, i.e. it looks the same in the bare and renormalised quantities and the renormalisation of the second term exseeds the accuracy of the two-loop  approximation. In fact, it gives a three-loop order term that is exactly what is needed to make the three-loop contribution finite, as will be seen later.

Thus, one can equally write eq.(\ref{twoloop}) in terms of renormalised quantities
\beq
W(\Phi_R)=\lambda_R \Phi_R^3+\frac{|\lambda_R|^4}{(4\pi)^4} \frac{1}{2}\zeta(3)~\lambda_R\Phi_R^3+O(|\lambda_R|^6)\lambda_R\Phi_R^3.\label{twoloop2}
\eeq
As we will see, a similar situation is partially the case in the three-loop approximation as well.

\section{Three-loop chiral correction}

Consider now the three-loop calculations of the chiral effective potential. The basic condition we should take into account  is  that the number of chiral and anti-chiral supercovariant derivatives differs by one and that the number of external chiral lines is equal to three as it is dictated by \eqref{cond}.
\begin{figure}[ht]
 \begin{center}
  \epsfxsize=12cm
 \epsffile{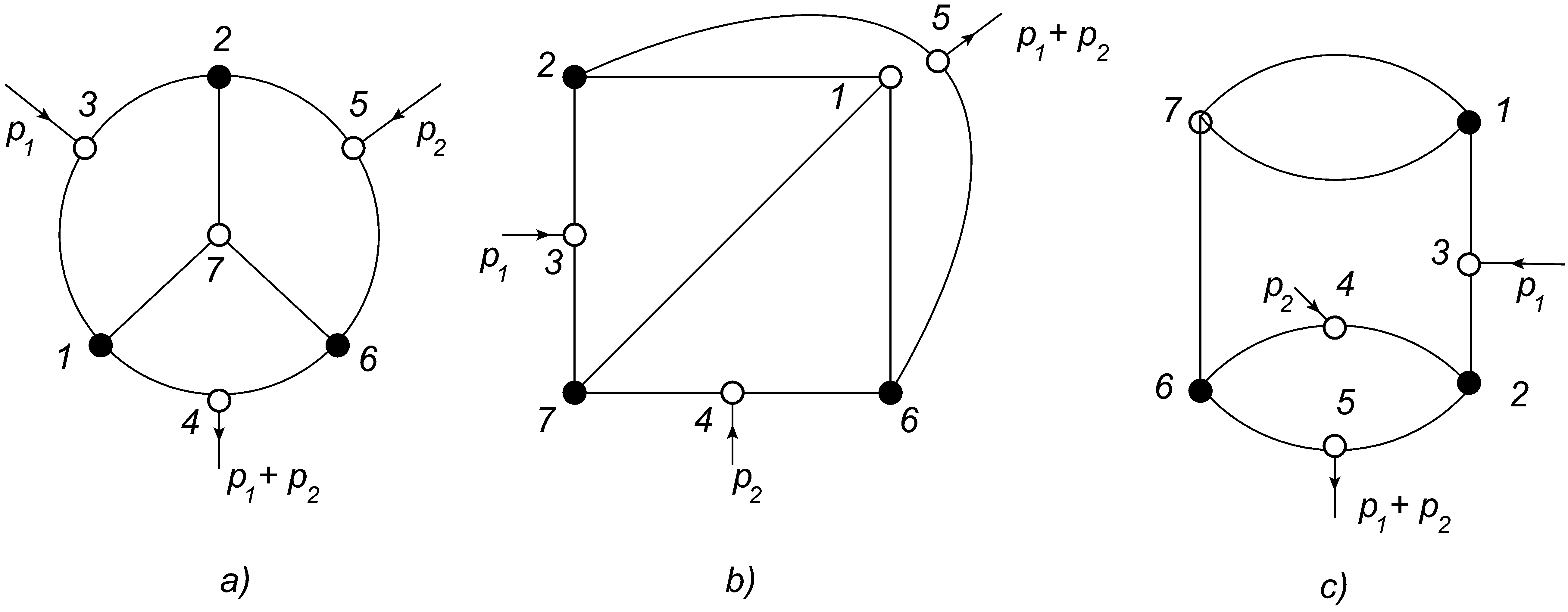}
 \end{center}
 \vspace{-0.2cm}
 \caption{Graphs contributing to the holomorphic effective potential at the three-loop level:  a) Mercedes-type diagram, b) non-planar triple loop diagram, c) barrel-type diagram}
\label{hthreeloop}
 \end{figure}
There are only three supergraphs that satisfy such requirements at the three-loop level. They are shown in Fig. \ref{hthreeloop}. Two of these diagrams (namely a) and b) ) are finite in terms by power counting and the last one c) is divergent, since it contains a one-loop divergent
subgraph of the propagator type. We calculate these diagrams one by one.

\subsection{Supergraph a)}

\begin{figure}[ht]
 \begin{center}
  \epsfxsize=13cm
 \epsffile{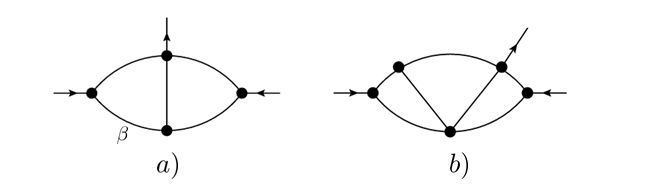}
 \end{center}
 \vspace{-0.2cm}
 \caption{The scalar master integral topologies giving a finite contribution to holomorphic potential.
 The contribution from barrel-type diagrams is proportional to a), meanwhile the non-planar and Mercedes-type
 diagrams giv rise to b) scalar graphs. Here $\beta$ can be  non-integer number. }
\label{topologies}
 \end{figure}

Let us first consider the diagram a). It is given by the Mercedes-type supergraph\footnote{Such a supergraph was also considered in the work \cite{West-1}, but the corresponding
momentum integral was not calculated in explicit form. }.
The corresponding contribution has the form
\beq
\begin{gathered}
\Gamma^{(3a),h}(\Phi)=\lambda\frac{|\lambda|^6}{3!\times 8}\int  \prod_{i=1}^7d^8z_i ~\Phi(z_3)\Phi(z_4)\Phi(z_5) \left\{ \frac{1}{\square_1}\delta_{1,3} \frac{\bar{D}_3^2 D_6^2}{16\square_6}\delta_{3,6} \frac{D_6^2 \bar{D}_5^2}{16\square_6}\delta_{6,5} \frac{D_2^2}{4\square_2} \delta_{5,2} \right. \\ \left.
\frac{1}{\square_2} \delta_{2,4}
\frac{\bar{D}_4^2 D_1^2 }{16\square_1} \delta_{4,1}
\frac{1}{\square_1} \delta_{1,7}
\frac{\bar{D}_7^2 D_6^2 }{16\square_6} \delta_{7,6}\frac{\bar{D}_7^2 D_2^2 }{16\square_2} \delta_{7,2}\right\}
\label{G3a}
\end{gathered}
\eeq
The loop integrand in momentum space can be expressed as
\beq
\begin{gathered}
I_3^{(a)}(p_1,p_2)= \lim_{p_1,p_2 \rightarrow0}\int \prod_{i=1}^3\frac{d^4 q_i}{(4\pi)^{4}}
\frac{(a_1 p_1^2+a_2 p_2^2 +a_3 (p_1 \cdot p_2))}{q_1^2 (q_1+p_1)^2 (p_1+q_2)^2 (q_2+p_1+p_2)^2 (q_2+p_1+p_2)}\times \\ \times\frac{1}{(q_3)^2 (q_1-q_2)^2 (q_1-q_3)^2 (q_3-q_2)^2}
\end{gathered}
\eeq
and the coefficients in the numerator are given
\beq
\begin{gathered}
a_1=q_1^2 \left(q_2-q_3\right)^2,   \\
a_2= q_2^2 \left(q_1-q_3\right)^2, \\
a_3=-2\left(q_3^2(q_1\cdot q_2) -q_2^2 (q_3\cdot q_1)+q_1^2 (q_2\cdot q_3)\right).
\end{gathered}
\eeq
As it was mentioned before, after calculation of the integral we should take the limit $p_i \rightarrow 0$.
Note that the reduced scalar master-integral has the topology depicted in Fig \ref{topologies}b) which is known to be evaluated by the uniqueness method \cite{Kazakov:1983pk, Usyukina:1994iw}. Note that the dimensionality of the diagrams in Fig. \ref{topologies} is proportional to the inverse square of the momentum $p_i$, but each of the diagrams is multiplied by the corresponding factor of squared $p_i$ derived from the numerator, so they cancel each other. 

The total contribution has the following form (hereafter we omit $4\pi$ factors in the results):
\beq
\label{a}
\begin{gathered}
\Gamma^{(3a),h}[\Phi]=|\lambda|^6 \frac{5}{12}\zeta(5)~\int d^6 z ~  \lambda \Phi^3(z)
\end{gathered}
\eeq
We stress that this expression is finite and does not require any renormalisation.

We note that this diagram belongs to a class of diagrams with a predictable iterative structure and reminds zig-zag-type diagrams, which have been extensively studied in the framework of fishnet models \cite{Derkachov:2022ytx}, and represents a special case of the superfishnet diagram arising from the double-scaling limit in  the $\beta$-deformed $\mathcal{N}=4$ super Yang-Mills model \cite{Kade:2024ucz}. We leave the discussion of possible applications of the fishnet formalism to this class of integrals for further works.

\subsection{Supergraph b)}

The next supergraph b) is non-planar. The corresponding contribution is given by the following
expression:
\beq
\begin{gathered}
\Gamma^{(3b),h}(\Phi)=\lambda\frac{|\lambda|^6}{3!\times 8}\int  \prod_{i=1}^7d^8z_i ~\Phi(z_3)\Phi(z_4)\Phi(z_5) \left\{
\frac{1}{\square_2}\delta_{2,3}\frac{\bar{D}^2_3 D^2_7}{16 \square_{7}}\delta_{3,7} \frac{\bar{D}^2_7 D^2_4}{16 \square_{7}}\delta_{7,4} \right. \\ \left. \frac{1}{\square_6}\delta_{4,6}\frac{D^2_6}{4 \square_6}\delta_{6,5} \frac{\bar{D}^2_{5}D^2_{2}}{16 \square_2} \delta_{5,2} \frac{D^2_{2}\bar{D}^2_{1}}{16 \square_2} \delta_{2,1} \frac{\bar{D}^2_{1} D^2_{6}}{16 \square_6} \delta_{1,6} \frac{1}{\square_7}\delta_{1,7}
\right\}
\label{G3b}
\end{gathered}
\eeq
After the $D$-algebra routine this integral is reduced to the following scalar integral:
\beq
\begin{gathered}
I_3^{(b)}(p_1,p_2)= \lim_{p_1,p_2 \rightarrow0}\int \prod_{i=1}^3\frac{d^4 q_i}{(4\pi)^{4}}   ~\frac{(b_1 p_1^2+b_2 p_2^2 +b_3 (p_1 \cdot p_2))}{(q_1-p_1)^2 q_1^2 q_2^2 (q_2+p_2)^2 (q_3+p_2)}\times \\ \times\frac{1}{(q_3-p_1)^2 (q_3-q_1)^2 (q_3-q_2)^2 (q_1-q_2)^2}
\end{gathered}
\eeq
where
\beq
\begin{gathered}
b_1=q_2^2\left(q_1-q_3\right)^2,   \\
b_2= q_1^2\left(q_2-q_3\right)^2, \\
b_3=-2\left(q_1^2 \left(q_2-q_3\right)^2+q_3^2\left(q_1-q_2\right)^2 -q_2^2 \left(q_1-q_3\right)^2\right).
\end{gathered}
\eeq
The calculations similar to the previous one give the result
\beq
\label{b}
\begin{gathered}
\Gamma^{(3b),h}[\Phi]= \frac{5}{12}\zeta(5)|\lambda|^6~ \int d^6 z ~\lambda\Phi^3(z)
\end{gathered}
\eeq
Like in the previous case, the result is finite and does not require any renormalisation.

\subsection{Supergraph c)}

At last,  we consider the supergraph c). The corresponding contribution is given by expression
\beq
\begin{gathered}
\Gamma^{(3c),h}(\Phi)=\lambda \frac{|\lambda|^6}{3! \times 8}\int  \prod_{i=1}^7d^8z_i ~\Phi(z_3)\Phi(z_4)\Phi(z_5) \left\{ \frac{1}{\square_1}\delta_{1,3} \frac{\bar{D}^2_3 D^2_2}{16\square_2}\delta_{3,2}\frac{\bar{D}^2_4 D^2_2}{16\square_2}\delta_{2,4} \frac{ D^2_6}{4\square_6} \delta_{4,6} \right. \\ \left. \frac{\bar{D}^2_5 D^2_6}{16\square_6}\delta_{6,5}  \frac{1}{\square_2}\delta_{5,2} \frac{1}{\square_6}\delta_{6,7}  \frac{\bar{D}^2_7 D^2_1}{16\square_1} \delta_{1,7}\frac{\bar{D}^2_7 D^1_2}{16\square_1}\delta_{7,1} \right\}
\label{G3c}
\end{gathered}
\eeq
Note that this integral contains a one-loop divergence so that the total integral is divergent. If the $D$-algebra routine results in terms proportional to any of propagators of this subgraph, the whole integral gives a zero contribution (tadpole diagrams are zero because there is no mass term). Thus, only numerators in the form of the scalar product of the corresponding internal momenta can contribute to the final result. Namely, if this loop is given by the integral
\beq
\int \frac{d^4q_3}{(4\pi)^4}\frac{c_1 ~q_3^2+c_2 ~(q_3-q_1)^2+ c_3 ~q_3 \cdot  l }{q_3^2 (q_3-q_1)^2}
\eeq
where $c_i$ are some coefficients fixed by a covariant derivatives permutation, only those proportional to $c_3$ should be taken into account.
The integrand of \eqref{G3c} after the $D$-algebra evaluation and Fourier transform can be represented in the following form in momentum space:
\beq
\begin{gathered}
\label{intC}
I_3^{(c)}(p_1,p_2)= \lim_{p_1,p_2 \rightarrow0}\int \prod_{i=1}^3\frac{d^4 q_i}{(4\pi)^{4}}  \frac{( c_1 p_1^2+c_2 p_2^2 +c_3 (p_1 \cdot p_2))}{q_3^2 (q_3-q_1)^2 q_1^4 (q_1+p_1)^2 (q_1-q_2)^2   }\times\\ \times\frac{1}{(q_2+p_1)^2(q_1-q_2+p_2)^2 (q_2-p_2)^2}
\end{gathered}
\eeq
where the $c_i$ are calculated as
\beq
\begin{gathered}
c_1=-q_1^2 (q_1-q_2)^2,   \\
c_2=-q_1^2 (q_1-q_2)^2, \\
c_3=-2q_1^2 (q_1-q_2)^2,
\end{gathered}
\eeq
with the help of the well-known formulas of integration (see e.g. \cite{Kazakov:1984mm}).

The result in the minimal subtraction scheme takes the form
\beq
\Gamma^{(3c),h}[\Phi]=-|\lambda|^6\left(\frac{3\zeta (3)}{8 \epsilon }+\frac{3}{2}\zeta (3)+\frac{9}{16}\zeta(4)\right)\int d^6 z ~\lambda\Phi^3(z) \label{c}
\eeq

\begin{figure}[ht]
 \begin{center}
  \epsfxsize=4cm
 \epsffile{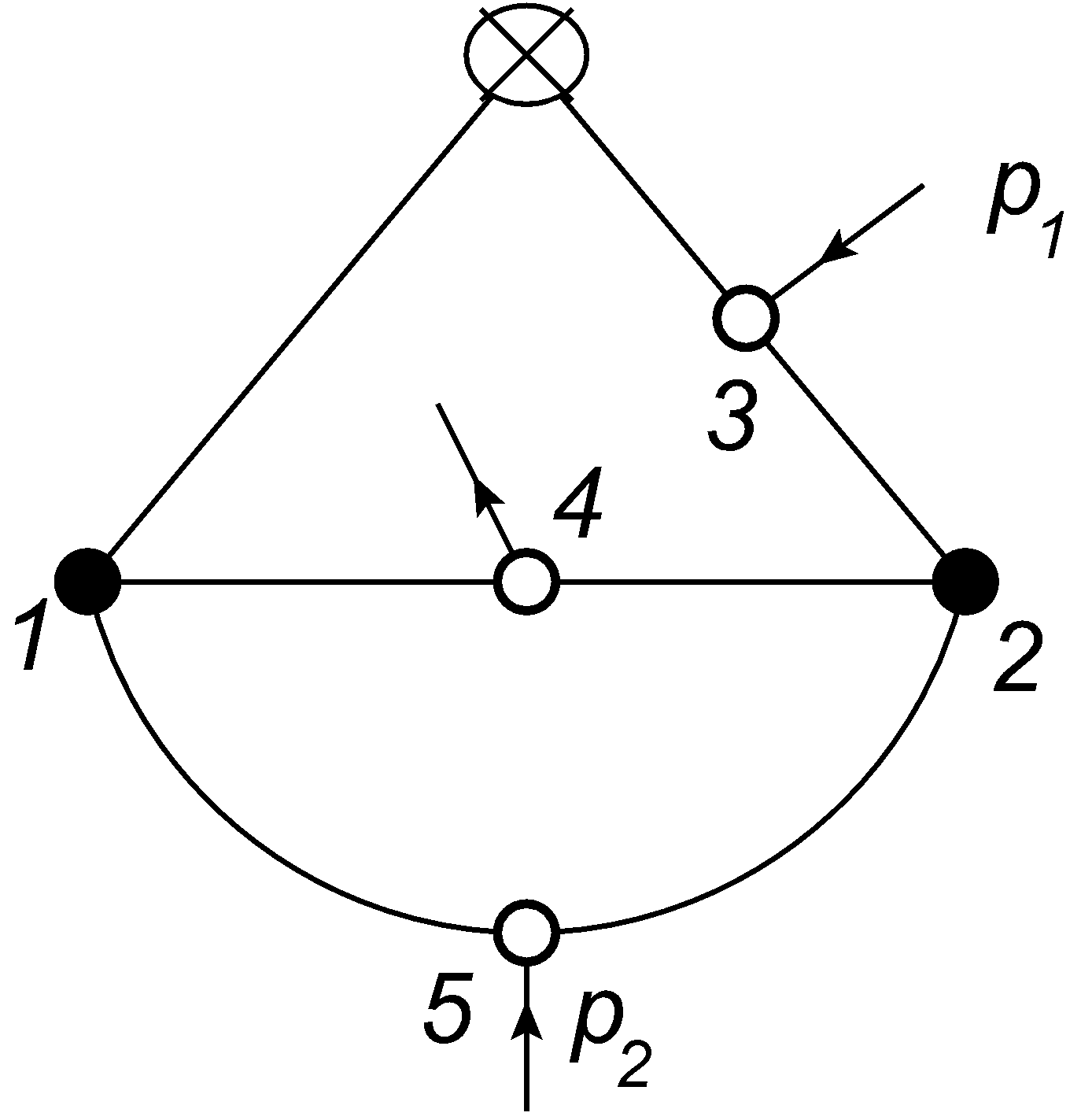}
 \end{center}
 \caption{The diagram with the counterterm}
\label{count}
 \end{figure}

As already mentioned, it contains the UV divergence. To remove it, one can act in two possible ways.
The first one is to subtract the divergence with the help of the one-loop counterterm
\beq
\label{counter}
\textbf{K}_{1,div}=\frac{|\lambda|^2}{2\epsilon}\Phi \bar{\Phi},
\eeq
 which gives the additional diagram shown in Fig.\ref{count}.
The other possibility is to write down the total expression first in terms of the bare quantities and then replace them by the renormalised ones according to eqs.(\ref{renorm}) with $z=1+\frac{|\lambda|^2}{2\epsilon}$. In what folows, we take the second way.

\subsection{Final result}

Summing up expressions (\ref{a}), (\ref{b}), (\ref{c})
(\ref{counter}),
we get the final expression for the three-loop chiral quantum contribution to the effective action
written in terms of the bare quantities in the form
\beq
\label{fin3}
\textbf{W}^{(3)}=-\frac{|\lambda|^6}{(4\pi)^8}\left(\frac{3\zeta (3)}{8 \epsilon }+
\frac{3}{2}\zeta(3)+\frac{9}{16}\zeta(4)-\frac{5}{6}\zeta(5)\right)~ {\lambda}\Phi^3(z).
\eeq
Summing the tree, two-loop and three-loop level results in an expression in the following form, one gets
\beq
W(\Phi)=\lambda \Phi^3\left[1+\frac{|\lambda|^4}{(4\pi)^4} \frac{1}{2}\zeta(3)-\frac{|\lambda|^6}{(4\pi)^8}\left(\frac{3\zeta (3)}{8 \epsilon }+
\frac{3}{2}\zeta(3)+\frac{9}{16}\zeta(4)-\frac{5}{6}\zeta(5)\right)+O(|\lambda|^8)\right]. \label{threeloop}
\eeq
Replacing now the bare fields and the bare coupling with the renormalised ones, according to eq.(\ref{renorm}),
we get rid of the UV divergence and get the renormalised expression
\beq
W_R(\Phi_R)=\lambda_R \Phi_R^3\left[1+\frac{|\lambda_R|^4}{(4\pi)^4} \frac{1}{2}\zeta(3)
-\frac{|\lambda_R|^6}{(4\pi)^8}\left(
\frac{3}{2}\zeta(3)+\frac{9}{16}\zeta(4)-\frac{5}{6}\zeta(5)\right)
+O(|\lambda_R|^8)\right].\label{threeloop2}
\eeq
The contribution to the effective action is then given by
\beq
\label{final-1}
\bar{\Gamma}_{chiral}[\Phi_R] = \int d^6z~ \textbf{W}_R(\Phi_R)
\eeq
with $W_R(\Phi_R)$ given by eq.(\ref{threeloop2}).
It is worth emphasizing that unlike generic tree-loop divergences, the result \eqref{threeloop} contains only one-loop divergences.

\section{Summary and discussion}

We have fulfilled  the three-loop calculations of the chiral effective potential in the massless Wess-Zumino model. The theory was formulated in terms of $\mathcal{N}=1$ superfields and studied in the framework of the corresponding supergraph technique. It was shown that there are only three of three-loop supergraphs giving rise to chiral effective potential. The calculation of the contributions of these supergraphs was given in Section 3 using the Wolfram Mathematica packages (e.g. \texttt{SusyMath.m}
\cite{Ferrari:2007sc}) and methods for evaluating multi-loop scalar master integrals.   An analogous at least four loop calculation seems quite possible.

We would like to stress ones more that the existence of the chiral effective potential indicates a violation of the Seiberg holomorphicity  in massless chiral-antichiral theories. The previous calculations \cite{West-1,Jack,West-2,Buch,BP} demonstrate the appearance of purely
chiral contributions to effective action at two loops. In the present paper, we show that it takes place in three loops as well. We expect that this effect is present in all loops.

It seems that the results of the present work can be extended to the the chiral-anti-chiral superfield models with
arbitrary classical Kähler and (anti)chiral superpotentials,
including the non-renormalisable ones. One may
hope to find restrictions on the non-renormalisable classical
potentials when quantum chiral corrections may turn out to be
finite.

Another interesting extension is to study a possibility to derive the chiral effective potential in the more general
chiral-antichiral models coupled to supersymmetric Yang-Mills
theories.  In particular, it may be interesting to compute the chiral quantum contributions to the effective action in
superconformal gauge theories including $\mathcal{N}=4$ SYM theory where one can hope to get non-perturbative results. In this regard, the ${\cal N}=2$ harmonic superspace methods \cite{GIOS} may be useful.

\appendix \section{Details of momentum integral calculation}\label{Appendix}

\begin{figure}[ht]
 \begin{center}
  \epsfxsize=13cm
 \epsffile{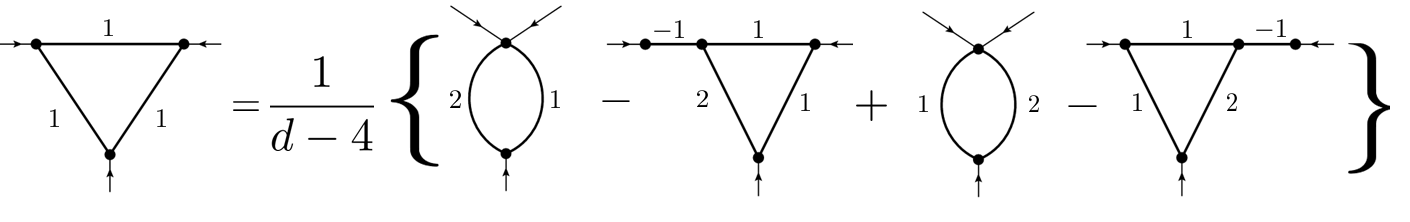}
 \end{center}
 \vspace{-0.1cm}
 \caption{Single momentum triangle Feynman diagram decomposition. The numbers on the lines denote the degree of the propagator}
\label{triangle}
 \end{figure}

\begin{figure}[ht]
 \begin{center}
  \epsfxsize=16cm
 \epsffile{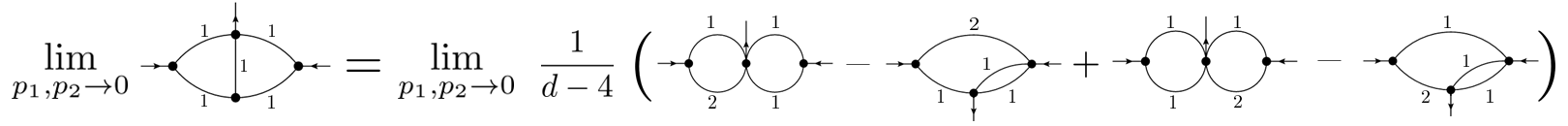}
 \end{center}
 \vspace{-0.1cm}
 \caption{Graphical representation of calculation of the two-loop scalar master integral according to integration by parts \eqref{twoloopmasterind3}.}
\label{walnut_c}
 \end{figure}

In this appendix we take a closer look at the computation of the two-loop diagram resulting from the $D$-algebra routine. 
Note that the initial two-loop supergraph is symmetric with respect to permutations of all three incoming momenta. Therefore, it is enough to divide the integrand of \eqref{twoloopmaster} by the summands of the numerator and to consider one of them to obtain the complete result, since the rest of the integrands can be integrated in the same way. Thus the integral to be calculated is as follows:
\beq
\begin{gathered}
I^{(1)}_2=\lim_{p_{1,2}\rightarrow0}\int \frac{d^4 q_1}{(4\pi)^4} \frac{d^4 q_2}{(4\pi)^4}  \frac{ p_1^2}{q_2^2 (q_1+q_2)^2 (q_1-p_1)^2 (q_2-p_2)^2 (q_1+q_2-p_1-p_2)^2}
\label{twoloopmasterind}
\end{gathered}
\eeq
It can be observed that this diagram has a "walnut" topology. It is more convenient to calculate these massless diagrams using the method of integration by parts (IBP). Let us rewrite the diagram omitting $p_1^2$ and using the dimensional regularisation
\beq
I^{(1)}_2=\lim_{p_{1,2}\rightarrow0}\int  \frac{d^d q_1}{(4\pi)^d} \frac{d^d q_2}{(4\pi)^d}  \frac{1}{q_1^2 (q_1-p_1)^2 q^2_2 (q_2-p_2)^2 (q_1+q_2)^2}
\eeq
where $d=4-2\epsilon$. Note that the full divergence of the following integral is equal to zero
\beq
\begin{gathered}
\int d^d q_1 d^d q_2  \frac{\partial}{\partial q_1^\mu} \left[ (q_1-q_2)^\mu \mathcal{I}^{(1)}_{2}\right]=0
\label{twoloopmasterind2}
\end{gathered}
\eeq
where
\beq
\mathcal{I}^{(1)}_{2}=\frac{1}{q_1^2 (q_1-p_1)^2 q^2_2 (q_2-p_2)^2 (q_1+q_2)^2}
\eeq
is the integrand of $I^{(1)}_2$. Taking the derivative leads us to the following expression:
\beq
\begin{gathered}
I^{(1)}_2=\frac{1}{d-4}\int d^d q_1 d^d q_2 \left((q_1-q_2)^2\left[\frac{1}{q_1^2}+\frac{1}{(q_1-p_1)^2}\right]-\left[\frac{q_2^2}{q_1^2}+\frac{(q_2-p_1)^2}{(q_1-p_1)^2}\right]\right)\mathcal{I}^{(1)}_2
\label{twoloopmasterind3}
\end{gathered}
\eeq
This transformation is equivalent to rewriting left three-point one-loop subgraph of \eqref{twoloopmasterind} in the way shown in Fig. \ref{triangle}.
Note that the last numerators in the bracket in this expression do not reduce the corresponding propagators.  However, the integral is taken in the limit $p_{1,2} \rightarrow 0$, so the expression is considerably simplified because the terms proportional to $p_1$ are supressed. That is why  the expression \eqref{twoloopmasterind3} is diagrammatically depicted in Fig. \ref{walnut_c}.
This leads us to a simple formula by which the diagram is reduced to a sum of graphs that contain only simple two-point one-loop functions, which are shown in Fig. \ref{walnut_c}.
The two-point one-loop integrals can be calculated using the next formula
\beq
J_{\alpha,\beta}(k)=\int \frac{d^d q}{(q-k)^{2\alpha}q^{2\beta}}=\frac{a(\alpha)a(\beta)}{a(d/2-\alpha-\beta)}(k^2)^{d/2-\alpha-\beta}
\label{bubble}
\eeq
where $a(\alpha)=\Gamma(d/2-\alpha)/\Gamma(\alpha)$ with the dimensional regularisation $d=4-2\epsilon$.
The final result is given as $6\zeta(3)$ restoring the result in the main part of the paper. This number coincides with the results of many other similar calculations \cite{Chetyrkin:1981qh,Kazakov:1984mm, Grozin:2012xi,Usyukina:1994iw}.

\begin{figure}[ht]
 \begin{center}
  \epsfxsize=16cm
 \epsffile{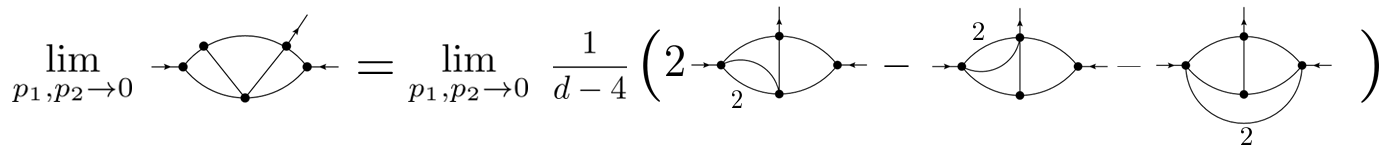}
 \end{center}
 \vspace{-0.1cm}
 \caption{Graphical representation of the calculation of the three-loop scalar master integral according to integration by parts. Unless otherwise noted, propagators in the diagram are regular.}
\label{3walnut_calc}
 \end{figure}

Once one has been able to establish the rules for integrating two-loop integrals, the integration of three-loop integrals can be accomplished in an absolutely identical step-by-step manner by integrating by parts of three-point subgraphs and integrating over a loop of two- and three-point subgraphs. For instance, after the $D$-algebra, the supergraph from Fig.\ref{hthreeloop}a is reduced to a scalar master integral of the form shown in Fig. \ref{topologies}b. In this scalar diagram, we first integrate the left three-point subdiagram by parts (integration corresponding to the transformation from Fig.\ref{triangle}) to obtain the sum of combinations of diagrams in the form of a two-loop graph calculated above with the known one-loop two-point diagram subgraphs \eqref{bubble}, this procedure is shown in Fig.\ref{3walnut_calc}.
Note that now one can see that the three-loop diagram is reduced to two-loop and one-loop diagrams we have already calculated. The remaining diagrams can be consistently integrated using the expression for the two-loop walnut \eqref{twoloopmasterind3} and one-loop \eqref{bubble} diagrams with the help of dimensional regularisation. The final result is obtained as $20\zeta(5)$. Also note that often calculations of scalar integrals like these are performed in the coordinate representation, in which the rules of integration have a more concise and universal form  (see \cite{Kazakov:1984mm,Usyukina:1994iw} for details of calculation).

\end{document}